\documentclass[10pt]{article}
\usepackage{amssymb,amsmath,mathrsfs,braket}
\usepackage{graphicx,rotate}
\usepackage{cite,wrapfig}

\usepackage{color}
\long\def\rpl#1!!#2!!{\textcolor{red}{#1} \textcolor{blue}{#2}}

\evensidemargin=\oddsidemargin
\def\Eqn#1{Eq.\ (\ref{#1})}

\def\mxi{m_{\xi^+}}
\def\tb{\tan\beta}

\begin{document}
\begin{flushright}
SINP/TNP/2014/02
\end{flushright}

\begin{center}
{\Large \bf Feasibility of light scalars in a class of two-Higgs-doublet models
 and their decay signatures} \\
\vspace*{5mm}  {\sf
    Gautam Bhattacharyya$^1$, Dipankar Das$^1$,  Anirban Kundu$^2$}

{\small } $^{1)}${\em Saha Institute of Nuclear Physics, 1/AF Bidhan
  Nagar, Kolkata 700064, India}
\\
\vspace{3pt} {\small } $^{2)}${\em Department of Physics, University of
Calcutta,\\
92 Acharya Prafulla Chandra Road, Kolkata 700009, India}
\normalsize
\end{center}

\begin{abstract}
We demonstrate that light charged and extra neutral scalars in the
(100-200) GeV mass range pass the potentially dangerous flavor
constraints in a particular class of two-Higgs-doublet model which has
appropriately suppressed flavor-changing neutral currents at tree
level. We study their decay branching ratios into various fermionic
final states and comment on the possibility of their detection in the
collider experiments. We also remark on how their trademark decay
signatures can be used to discriminate them from the light nonstandard
scalars predicted in other two-Higgs-doublet models.

\end{abstract}


\section{Introduction}
Plenty of motivations exist for considering the recently discovered
Higgs boson at the CERN Large Hadron Collider (LHC) to be a member of
a richer scalar structure beyond what is predicted by the Standard
Model (SM). An exciting possibility is the two-Higgs-doublet model
(2HDM) \cite{Branco:PhysRept}, which receives special attention
because the minimal supersymmetric standard model relies on
it. Extremely tight experimental constraints on tree level
flavor-changing neutral currents (FCNC) led to the
Glashow-Weinberg-Paschos theorem in the multi-Higgs
context\cite{Glashow:1976nt,Paschos:1976ay}. This implies that the
absence of tree level FCNC will be ensured if all right-handed
fermions of a given charge couple to a single Higgs doublet. In 2HDM,
this can be achieved by the introduction of discrete or continuous
symmetries that act on the scalars and fermions. Following the $Z_2$
discrete symmetry under which $\Phi_1 \to - \Phi_1$ and $\Phi_2 \to
\Phi_2$, four types of 2HDM emerge, based on the fermion
transformations under that symmetry.  They are classified as (i) Type
I: all quarks and leptons couple to only one scalar $\Phi_2$; (ii)
Type II: $\Phi_2$ couples to up-type quarks, while $\Phi_1$ couples to
down-type quarks and charged leptons (minimal supersymmetry conforms
to this category); (iii) Type Y (or III, or flipped): $\Phi_2$ couples
to up-type quarks and leptons, while $\Phi_1$ couples to down-type
quarks; (iv) Type X (or IV, or lepton specific): $\Phi_2$ couples to
all quarks, while $\Phi_1$ couples to all leptons.  Experimental
constraint on the charged Higgs mass in the Type II and Y models is
quite strong: $\mxi > 300$ GeV for any $\tb\equiv v_2/v_1$, which is
the ratio of the two vacuum expectation values. This arises mainly
from the $b \to s \gamma$ constraint, but this constraint is
considerably weak in Type I and X scenarios \cite{Mahmoudi:2009zx}.

In this paper, we consider a special category of 2HDM formulated by
Branco, Grimus and Lavoura (hereafter called the BGL scenario)
\cite{Branco:1996bq}, where tree level FCNC exists with appropriate
suppression arising from the elements of the Cabibbo-Kobayashi-Maskawa
(CKM) matrix\footnote{A nonrenormalizable version of a similar
  scenario was constructed in \cite{Hadeed:1985xn}.}. Unlike the
general 2HDM with tree level FCNC
\cite{Crivellin:2013wna,Atwood:1996vj}, the BGL models introduce no
new parameters in the Yukawa sector, and therefore, are more
predictive.  In this scenario, instead of the discrete $Z_2$ symmetry
a global U(1) symmetry acts on a particular generation $i$ at a time,
as follows:
\begin{eqnarray}
Q_{Li}\to e^{i\theta}Q_{Li}\,,~~u'_{Ri}\to
e^{2i\theta} u'_{Ri}\,,~~\Phi_2\to e^{i\theta}\Phi_2 \,.
\label{BGL symmetry}
\end{eqnarray}
Here $Q_{Li}=(u'_{Li}\,,d'_{Li})^T$ is the left-handed quark doublet
for the $i$-th generation ($i=1,2,3$), while $u'_R$ denotes the
up-type right-handed quark singlets, all in the weak basis.  The
scalar doublet $\Phi_1$ and the other quark fields remain unaffected
by this transformation.  For this particular choice of the symmetry,
there will be no FCNC in the up sector and the FCNC in the down sector
will be controlled by the $i$-th row of the CKM matrix. This will lead
to three variants which will be called u-, c- and t-type models
according to $i=$ 1, 2, and 3 respectively\footnote{The other three
  variants can be obtained by replacing $u'_R$ in Eq.\ (\ref{BGL
    symmetry}) with $d'_R$ (down-type singlet), as a result of which
  there will be no FCNC in the down sector and the FCNC in the up
  sector will be controlled by the columns of the CKM matrix. We do
  not consider this scenario primarily because the FCNC in the up
  sector is less restrictive.}.  A number of low-energy observables
severely constrain the u- and c-type models, as we will show later; so
it makes sense to talk about the t-type model only. We will also show
from the same observables that in the t-type model one can entertain
charged Higgs mass in the ballpark of 150 GeV for $\tb > 1$. In fact,
the additional scalars, namely, the CP-even $H$ and the CP-odd $A$
together with the charged scalar $\xi^+$, could all be taken in the
100-200 GeV mass range.  These light scalars would leave distinct
decay signatures in the collider experiments. By comparing their
branching ratios in various flavor channels, it is possible to
distinguish the BGL scalars from the light ones predicted in Type I
and X models. Our study assumes special significance in view of the 14
TeV run of the LHC, its possible upgrade to higher luminosity, and the
possibility of precision studies at the International Linear Collider
(ILC).

The paper is arranged as follows. In Section II, we briefly introduce
the BGL model. Section III deals with the low-energy constraints on
the model.  In particular, we show that the BGL model allows a light
charged Higgs. We also point out the differences between BGL and other
2HDMs. The decay patterns and possible signatures of the nonstandard
scalars are discussed in Section IV. We summarize and conclude in the
last section.  The appendices contain all the relevant formulae that
have gone into our calculations.

\section{Yukawa sector of 2HDM (BGL)}
The scalar potential of the BGL model is identical to the other
canonical 2HDMs.  This scenario is manifestly CP conserving.  Let us
denote the CP-even neutral components of $\Phi_1$ and $\Phi_2$ by
$\rho_1/\sqrt{2}$ and $\rho_2/\sqrt{2}$ respectively, with $\langle
\rho_{1(2)}\rangle = v_{1(2)}$. They are of course weak
eigenstates. The corresponding mass eigenstates $H$ and $h$ can be
obtained by diagonalizing the $2\times 2$ mass matrix ${\cal M}$ using
an orthogonal transformation characterized by an angle $\alpha$, given
by
\begin{equation}
\cos 2\alpha \equiv \frac{ {\cal M}_{11} - {\cal M}_{22} } 
{ \sqrt{ ({\cal M}_{11} - {\cal M}_{22})^2 + 4 {\cal M}_{12}^2 } }\,.
\label{alpha-def}
\end{equation}
By convention, we will take $m_h < m_H$, and $h$ to be the 125 GeV scalar 
resonance discovered at the LHC. 

To study the Yukawa sector of the BGL model, it is helpful to go to
another neutral scalar basis $\{H^0,R\}$. This is not the mass basis
in general, and is obtained from the $\{\rho_1,\rho_2\}$ basis by a
rotation through the angle $\beta \equiv \tan^{-1}(v_2/v_1)$.  The
same rotation picks out the physical charged Higgs ($\xi^+$) and
charged Goldstone ($G^+$), as well as the physical pseudoscalar ($A$)
and neutral Goldstone ($G^0$), from their respective weak basis
states. In the CP-even neutral sector, $\langle H^0 \rangle =
v=\sqrt{v_1^2+v_2^2} = 246$ GeV, while $\langle R \rangle = 0$. The
relationship between the two bases $\{H,h\}$ and $\{H^0,R\}$ is given
by
\begin{subequations}
\begin{eqnarray}
H^0 &=& \cos(\beta-\alpha)H + \sin(\beta-\alpha)h \,, \\
R &=& -\sin(\beta-\alpha)H + \cos(\beta-\alpha)h \,.
\end{eqnarray}
\end{subequations}
The $H^0$ state has gauge and Yukawa couplings identical to those of
the SM Higgs boson.  The physical state $h$, observed at the LHC,
conforms to the state $H^0$ in the decoupling limit $\beta = \alpha
\pm \pi/2$ \cite{Gunion:2002zf}.  This decoupling limit in the 2HDM
context is now being increasingly motivated by the LHC data
\cite{Eberhardt:2013uba,Craig:2013hca}.

The Yukawa Lagrangian of the BGL model, worked out in
\cite{Branco:1996bq}, is given by
\begin{eqnarray}
{\cal L}_{Y} &=&
-\frac{1}{v}{H^0}\left[{\bar{d}}D_d{d}+{\bar{u}}D_u{u}\right]
+ \frac{1}{v}{R}\left[{\bar{d}}(N_dP_R+N_d^\dagger
  P_L){d}+{\bar{u}}(N_uP_R+N_u^\dagger P_L){u}\right] \nonumber \\
&& +\frac{i}{v}{A}\left[{\bar{d}}(N_dP_R-N_d^\dagger P_L){d}-{\bar{u}}
(N_uP_R-N_u^\dagger P_L){u}\right] 
+\left\{\frac{\sqrt{2}}{v}\xi^+ \bar{u}\left(VN_dP_R-N_u^\dagger 
VP_L\right){d} + {\rm h.c.}\right\} \,.
\label{BGL Yukawa}
\end{eqnarray}
Here, $u$ and $d$ stand for 3-generation up and down quarks in mass
basis, $D_u$ and $D_d$ are diagonal up and down mass matrices, and $V$
is the CKM matrix.  The matrices $N_u$ and $N_d$, for the u-, c- and
t-type models, have the following form (the $(i,j)$ indices in $N_d$
refer to $(d,s,b)$ quarks and the superscripts in bold font refer to
the model type):
\begin{subequations}
\begin{eqnarray}
N_u^{\bf u} &=& {\rm diag}\{-m_u \cot\beta\,, m_c\tan\beta\,,
m_t\tan\beta\}\,, \ \ \ 
(N_d)_{ij}^{\bf u} = \tan\beta~m_i \delta_{ij}
-(\tan\beta+\cot\beta)V_{ui}^* V_{uj} m_j\, , \\
N_u^{\bf c} &=& {\rm diag}\{m_u \tan\beta\,, -m_c\cot\beta\,,
m_t\tan\beta \}\,, \ \ \ 
(N_d)_{ij}^{\bf c} = \tan\beta~m_i \delta_{ij}
-(\tan\beta+\cot\beta)V_{ci}^* V_{cj} m_j\, , \\
N_u^{\bf t} &=& {\rm diag}\{m_u \tan\beta\,, m_c\tan\beta\,,
-m_t\cot\beta \}\,, \ \ \ 
(N_d)_{ij}^{\bf t} = \tan\beta~m_i \delta_{ij}
-(\tan\beta+\cot\beta)V_{ti}^* V_{tj} m_j\,.
\end{eqnarray}
\label{nund}
\end{subequations}
In the leptonic sector (with only left-handed neutrinos), the Yukawa
couplings of \Eqn{BGL Yukawa} should be read with the replacement
$(N_u, D_u)\to 0$, $V=1$, and $N_d(D_d)\to N_e(D_e)$, with $N_e$
resembling the diagonal $N_u$ matrices in Eq.~(\ref{nund}) with
appropriate replacement of quark masses by the charged lepton masses.
This means that there is no FCNC in the leptonic sector when the
neutrinos are considered to be massless.

\begin{figure}[!htbp]
\includegraphics[scale=0.4]{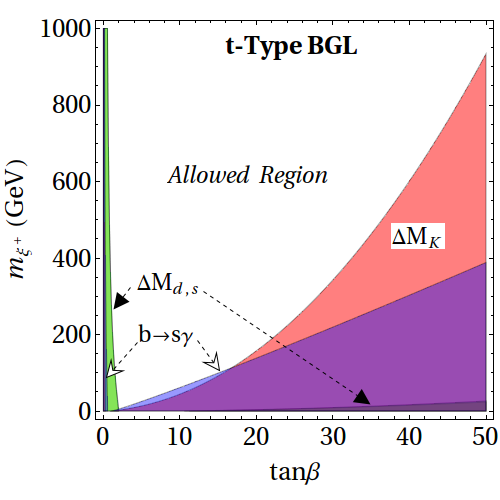}
~~~~~~~~~ 
\includegraphics[scale=0.4]{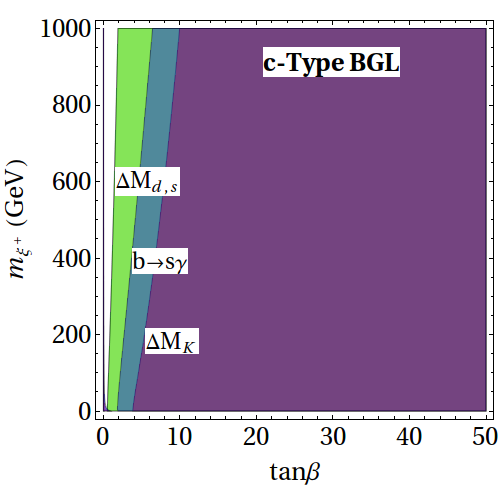}
\caption{\em Constraints from various observables for t- and
  c-(u-) type BGL models. In the left panel (t-type), for large
  $\tan\beta$, $\Delta M_K$ offers a stronger constraint than $b \to
  s \gamma$. The vertical spiked shaded region in the extreme
  left also correspond to the entire disallowed region in Type I and X
  models. In the right panel (c- or u-types), $\Delta M_d$ and $\Delta M_s$
  provide the most stringent constraints. Note that an assumption $m_H
= m_A$ has been made to switch off the tree level contribution to
the neutral meson mass differences.}
\label{tmodel1}
\end{figure}

The CP-odd scalar mass eigenstate $A$ would be massless if the
symmetry of \Eqn{BGL symmetry} is exact in the Higgs potential. Thus,
in the 't~Hooft sense, a light pseudoscalar will be natural in these
models.  While there are five free parameters in any BGL model,
namely, $\alpha$, $\beta$, $\mxi$, $m_H$, and $m_A$, we can make some
reasonable simplifications.  Considerations of perturbativity and
stability of scalar potential ensure that $m_A \sim m_H$ if $\tan\beta
\geq 10$ \cite{Bhattacharyya:2013rya}. If $m_A$ and $m_H$ are large,
we can even bring down the $\tan\beta$ limit further, say up to
$\tan\beta = 5$. However, for the sake of simplicity and economy of
parameters, we will assume $m_H = m_A$ for the remainder of this paper
unless explicitly mentioned otherwise.  Thus, in the decoupling limit,
i.e. $\cos(\beta - \alpha) = 0$, we are left with only three unknown
parameters: $\tan\beta$, $\mxi$ and $m_{H/A}$. It should be noted
though that consistency with the oblique $T$-parameter requires $\mxi
\sim m_H$ once we assume $m_H = m_A$ \cite{Bhattacharyya:2013rya}.

\section{Constraints on the parameter space}
\subsection{Neutral meson mixing}
Neutral meson mass
differences offer important constraints. 
The tree-level scalar exchange contribution to the
off-diagonal element of the $2\times 2$ Hamiltonian matrix is given by
\cite{Branco:1996bq}
\begin{equation}
(M_{12}^K)^{\rm BGL} \approx 
\frac{5}{24}\frac{f_K^2 m_K^3}{v^2} (V_{id}^* V_{is})^2 \frac{1}{{\cal
    A}^2} \,,
\label{m12}
\end{equation}
where $m_K$ is the neutral kaon mass and $f_K$ is the decay constant.
Similar expressions exist for $B_d$ and $B_s$ systems.
 The mass difference is given by $\Delta M_K \approx 2 \vert M_{12}^K\vert$. 
The contributions of three neutral scalars are contained in
\begin{equation}
\frac{1}{{\cal A}^2} = (\tan\beta +\cot\beta)^2\left( \frac{\cos^2(\beta-\alpha)}{m_h^2}
+\frac{\sin^2(\beta-\alpha)}{m_H^2} -\frac{1}{m_A^2} \right) 
= (\tan\beta +\cot\beta)^2  \left(\frac{1}{m_H^2} -
\frac{1}{m_A^2}\right)\,.
\label{onebya2}
\end{equation}
The last equality in Eq.~(\ref{onebya2}) holds in the decoupling
limit.  The size of the prefactors in Eq.~(\ref{m12}) tells us that
$m_A=m_H$ is very well motivated from the neutral kaon mass difference
for the u- and c-type models.  For the t-type model, however, this
degeneracy is more of an assumption than a requirement especially for
$\tan\beta \sim 1$.

With the assumption $m_H = m_A$, the dominant contributions to neutral
meson mass differences come from the charged Higgs box diagrams. The
expressions for the loop-induced amplitudes are given explicitly in
Appendix A. In Fig.~\ref{tmodel1}, constraints have been placed
assuming that the new physics contributions saturate the experimental
values of $\Delta M$ \cite{Beringer:1900zz}.  For $\tb > 1$, $\Delta
M_d$ and $\Delta M_s$ severely restrict the u- and c-type models,
whereas the t-type model can admit a light charged Higgs, at least for
$m_H = m_A$. For large $\tb$, $\Delta M_K$ offers a stronger
constraint than $b \to s \gamma$ (discussed later) in the t-type model
due to the dominance of the charm-induced box graph.

\subsection{$b\to s\gamma$}
The process $b\to s\gamma$ offers severe constraint on the charged
Higgs mass \cite{Grinstein:1987pu,Borzumati:1998tg}.  For Type II and
Y models, in the charged Higgs Yukawa interaction, the up-type Yukawa
coupling is multiplied by $\cot\beta$ while the down-type Yukawa is
multiplied by $\tan\beta$.  Their product is responsible for setting
$\tb$-independent limit $\mxi > 300$ GeV for $\tan\beta>1$
\cite{Deschamps:2009rh,Mahmoudi:2009zx,Cheng:2014ova}.  In Type I and
X models, each of these couplings picks up a $\cot\beta$ factor, which
is why there is essentially no bound on charged Higgs mass for
$\tan\beta>1$ in these models \cite{Mahmoudi:2009zx}.

In the BGL class of models, the constraint on $\mxi$ is different from
that in Type I or Type X 2HDM (detailed expressions are displayed in
Appendix B). This is because the BGL symmetry of \Eqn{BGL symmetry}
does not respect family universality.  For the $i$-type BGL model, the
relevant Yukawa couplings contain an overall factor of $(-\cot\beta)$
for vertices involving the $i$-th generation up-type fermion and a
factor of $\tan\beta$ for the others. Consequently, the top loop
contribution to the $b\to s\gamma$ amplitude will grow as
$\tan^2\beta$ for u- and c-type models resulting in very tight
constraints on $m_\xi$ for $\tan\beta>1$. On the contrary, for t-type
models, the top-loop contribution will decrease with increasing
$\tan\beta$ and will hardly leave any effect for $\tan\beta>1$,
similar to what happens in the Type I and X models.  But unlike in the
latter scenarios, the charm loop amplitude in t-type BGL grows as
$\tan^2\beta$.  It becomes numerically important for large $\tb$ and
does not allow $\xi^+$ to be very light.

\begin{figure}[!htbp]
\begin{center}
\includegraphics[scale=0.4]{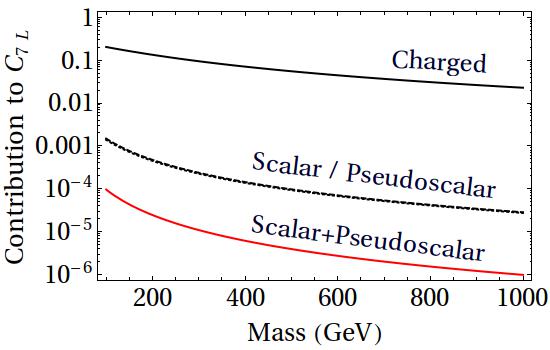}~~~
\includegraphics[scale=0.4]{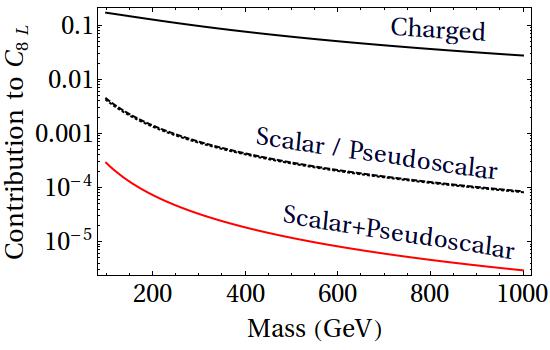}
\caption{\em Magnitude of the contributions to the effective Wilson
  coefficients $C_{7L}$ and $C_{8L}$ for $b\to s\gamma$, coming from
  $\xi^+$, $H$, and $A$, plotted against the corresponding masses. The
  middle curve in each panel shows the magnitude of the individual
  scalar and pseudoscalar contributions; they are too close to be
  differentiated in the shown scale. The lowest curve in each panel
  shows the sum of $H$ and $A$ contributions for the case $m_H = m_A$,
  which shows that the scalar and pseudoscalar contributions interfere
  destructively. $C_{7R}$ and $C_{8R}$ are suppressed by $m_s/m_b$ and
  are not shown here.}
\label{fig:c78}
  \end{center}
\end{figure}

Taking the branching ratio ${\rm Br}(b \to s \gamma) _{\rm SM}
=(3.15\pm 0.23)\times 10^{-4}$
\cite{Gambino:2001ew,Misiak:2006zs,Misiak:2006ab} and ${\rm Br}(b \to
s \gamma)_{\rm exp} = (3.55\pm 0.26)\times 10^{-4}$
\cite{Amhis:2012bh}, these features of the BGL models have been
displayed in Fig.~\ref{tmodel1}. The regions excluded at 95\% CL from
$b\to s\gamma$ have been shaded and appropriately marked.  Note that
we have considered not only the contributions from $(\xi^+, u_i)$
loops, but also from $(H/A, d_i)$ loops (due to tree level FCNC
couplings of $H$ and $A$). The numerical effects of the latter are
found to be small; we refer the reader to Fig.\ \ref{fig:c78}, where
separate contributions from the charged and the neutral scalars to
$C_{7L}$ and $C_{8L}$ are shown. The behaviour can also be intuitively
understood from the following comparison of the dominant contributions
from the charged and neutral scalar induced loops to the $b \to s
\gamma$ amplitude.  The ratio of $\xi^+$ and $(H/A)$-induced loop
\begin{wrapfigure}{r}{0.42\textwidth}
\includegraphics[scale=0.38]{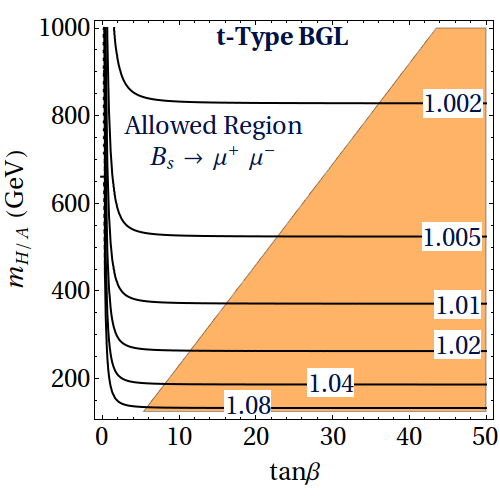}
\caption{\em The shaded region is disallowed by $B_s \to \mu^+ \mu^-$
  at 95\% CL. Contours of enhancements in $B_s\to\tau^+\tau^-$ over
  the SM estimate are also shown.}
  \label{fig:dilepton}
\end{wrapfigure}
contributions roughly goes like $(m_c^2 \tan^2\beta / m_b m_s)$ for
large $\tan\beta$, and $(m_t^2 \cot^2 \beta / m_b m_s)$ for
$\tan\beta$ of the order of one.  This justifies that the constraint
from $b \to s \gamma$ essentially applies on the charged Higgs mass.
In other words, that $\xi^+$ can be really light does not crucially
depend on the values of $m_H$ and $m_A$.  From now on, we stick only
to the t-type model to promote light charged Higgs phenomenology.

\subsection{Other constraints}
For t-type model, the branching ratios ${\rm Br}(B \to D^{(*)}
\tau\nu)$ and ${\rm Br}(B^+\to\tau\nu)$ do not receive any appreciable
contributions unless the charged Higgs mass is unnaturally small
defying the LEP2 direct search limit of 80 GeV
\cite{Searches:2001ac}. The process $B_s\to \ell^+\ell^-$ proceeds at
the tree level mediated by $H/A$ providing important constraints. The
amplitudes are proportional to $(\tan^2\beta+1)/m_{H/A}^2$ for $\ell =
e,\mu$, and $(\cot^2\beta+1)/m_{H/A}^2$ for $\ell=\tau$.  In
Fig.~\ref{fig:dilepton} we have shaded the region excluded at 95\% CL,
obtained by comparing the SM expectation of ${\rm Br}
(B_s\to\mu^+\mu^-) = (3.65\pm 0.23) \times 10^{-9}$
\cite{Bobeth:2013uxa} with its experimental value $(3.2\pm 1.0)\times
10^{-9}$ \cite{HFAGbs}. The details are provided in Appendix C. In the
same plot we display different contours for ${\rm
  Br}(B_s\to\tau^+\tau^-)/{\rm Br}(B_s\to\tau^+\tau^-)_{\rm SM}$,
where we observe slight enhancement over the SM expectation.

\section {Charged and neutral scalar branching ratios}

\begin{figure}[!htbp]
\includegraphics[scale=0.4]{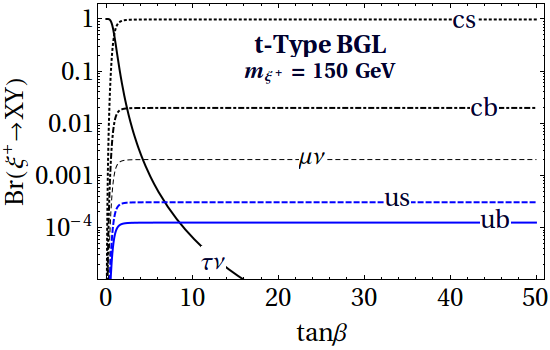}
~~~~~~~~~ 
\includegraphics[scale=0.4]{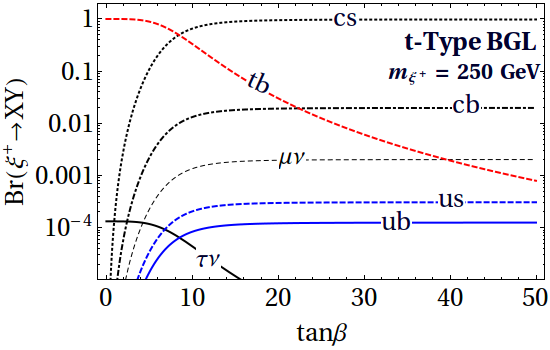}
\caption{\em The charged Higgs branching ratios to two-body final
  states for two benchmark choices of $\mxi$. }
\label{charged-br}
\end{figure}

In Type I model, the light charged Higgs goes to $\tau\nu$ and $cs$
(below the $tb$ threshold), and the branching ratios are independent
of $\tan\beta$, because both the leptonic and the quark couplings have
the same $\cot\beta$ prefactor \cite{Aoki:2011wd,Aoki:2009ha}. In Type
X model, the leptonic part has an overall $\tan\beta$ multiplicative
factor, so the charged Higgs preferentially decays into third
generation leptonic channels for large $\tan\beta$ (e.g. almost
entirely so for $\tb \geq 2.5$).  In the t-type BGL scenario, the
charged Higgs branching ratios into two-body fermionic final states
have been plotted in Fig.~\ref{charged-br}.  We have considered two
benchmark values for $\mxi$, one below the $tb$ threshold and the
other well above it. To a good approximation it is enough to consider
fermionic final states, because in the decoupling limit the $W^\pm
h\xi^\mp$ coupling vanishes and if we consider near degeneracy of
$\mxi$ and $m_{H/A}$ to satisfy the $T$-parameter constraint, then
$\xi^+$ cannot decay into $W^+ S^0$ ($S^0=H,A$) channel. Two
noteworthy features which distinguish the t-type BGL model from others
are: ($i$) the $\mu\nu$ final state dominates over $\tau\nu$ for
$\tan\beta > 5$, which is a distinctive characteristic of t-type BGL
model unlike any of the Type I, II, X or Y models (due to family
nonuniversal BGL Yukawa couplings); ($ii$) for $\tb > 10$, the
branching ratio into $cs$ significantly dominates over other channels
including $tb$, again a unique feature of t-type BGL.  The reason for
the latter can be traced to the relative size of the top and charm
quark masses {\em vis-\`{a}-vis} the $\tan\beta$ or $\cot\beta$
prefactor.  This will result in a dijet final state at the LHC,
without any $b$-jet, and hence the signal will be extremely difficult
to be deciphered over the standard QCD background.

\begin{figure}[!htbp]
\includegraphics[width=5.5cm,height=5cm]{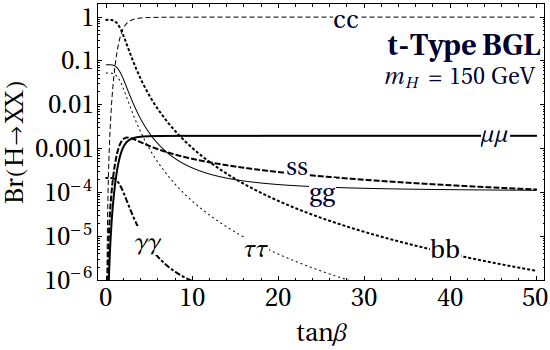}
\includegraphics[width=5.5cm,height=5cm]{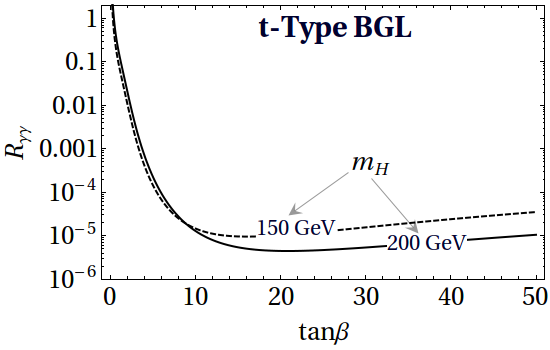}
\includegraphics[width=5.5cm,height=5cm]{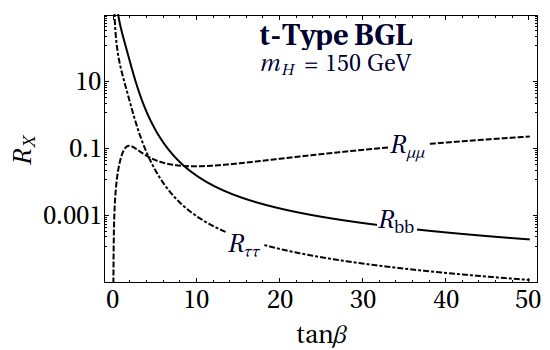}
\caption{\em For various two-body final states, the $R$ values, and the
  branching ratios of $H$.}
\label{neutral-br}
\end{figure}

We now discuss the decay branching ratios of the neutral scalar
$H$. In the decoupling limit $HVV$ ($V=W,Z$) coupling vanishes. Hence
we discuss flavor diagonal $ff$ final states (flavor violating modes
are CKM suppressed), together with $\gamma\gamma$ and $gg$ final
states.  In other types of 2HDM, the $bb$ and $\tau\tau$ final states
dominate over $cc$ and $\mu\mu$ channels, respectively
\cite{Aoki:2009ha}. Here, the hierarchy is reversed, which transpires
from the expressions of $N_d$ and $N_u$ in \Eqn{nund}.  To provide an
intuitive estimate of the signal strength, we define the following
variable:
\begin{eqnarray}
R_X = \frac{\sigma(pp\to H \to X)}{\sigma(pp\to h \to
  \gamma\gamma)} \, , 
\label{rx}
\end{eqnarray}
where the normalization has been done with respect to the SM Higgs
production and its diphoton decay branching ratio.  We recall that the
loop contributions of charged scalars to $h\to \gamma\gamma$ is tiny
as long as $m_{H/A} \simeq \mxi$ \cite{Bhattacharyya:2013rya}.  The
relative merits of various channels have been plotted in
Fig.~\ref{neutral-br}. The crucial thing to observe is that although
for $\tan\beta > 5$, $H$ decays entirely into dijet ($cc$), the
$\mu^+\mu^-$ mode may serve as a viable detection channel for $H$ in
future. With 20 fb$^{-1}$ luminosity at LHC8, the expected number of
diphoton events from the SM Higgs decay is about
400. Fig.~\ref{neutral-br} shows that $R_{\mu\mu} \sim 0.1$,
i.e. about 40 dimuon events from $H$ decay should have been
observed. However, they are going to be swamped by huge background
(mainly Drell-Yan, also QCD jets faking dimuon) \cite{cms-bkg}. At
LHC14 with an integrated luminosity of 300 fb$^{-1}$, we expect about
39000 $h\to\gamma\gamma$ events \cite{cerntwiki}, which means about
3900 $H\to\mu\mu$ events for $m_H=150$ GeV. Dimuon background studies
at 14 TeV are not yet publicly available. A rough conservative
extrapolation of the existing 7 and 8 TeV studies of the dimuon
background \cite{cms-bkg} gives us hope that the signal can be
deciphered over the background.  Note that these are all crude
estimates, made mainly to get our experimental colleagues interested
in probing such exotic decay modes.  A more careful study including,
e.g.  detection efficiencies and detailed background estimates, is
beyond the scope of this paper.  We emphasize that our scenario does
not say that $H, A$ or $\xi^+$ have to be necessarily light. If they
are heavy as they are forced to be in many other 2HDMs ($\sim 500$ GeV
or more), their direct detection in early LHC14 would be that much
difficult. The feature that makes our scenario unique is the {\em
  possibility} of their relative lightness as well as unconventional
decay signatures.

\section{Conclusions} 
We have shown that a particular class of two-Higgs-doublet model
admits charged and additional neutral scalars which can be as light as
$\sim$ 150 GeV. They successfully negotiate the stringent constraints
from radiative $b$-decay, neutral meson mass differences, and dimuon
decays of $B$ mesons. Special features of Yukawa couplings in this
model lead to characteristic decay signatures of the nonstandard
scalars, which are different from the signatures of similar scalars in
other 2HDM variants.  Preferential decays of both the charged and
additional neutral scalars into {\em second}, rather than the {\em
  third}, generation fermions for $\tb > 5$ constitute the trademark
distinguishing feature of this scenario, which can be tested in the
high luminosity option of the LHC or at the ILC.

\noindent {\em Note added:}~ During the finalization of this
manuscript, the paper \cite{Botella:2014ska} appeared which also deals
with the BGL scenario. We agree with their overall conclusion on the
feasibility of light charged Higgs boson. We have, however,
additionally analyzed the decay signatures of the new scalars.

\section{Acknowledgements}

We thank Sunanda Banerjee and
Swagata Mukherjee for helpful discussions.  AK acknowledges Department
of Science and Technology, Govt.\ of India, and Council for Scientific
and Industrial Research, Govt.\ of India, for research support. DD
thanks the Department of Atomic Energy, India, for financial support.

\appendix
\numberwithin{equation}{section}

\section{Neutral meson mixing}
The dominant one-loop effective Lagrangian for $\Delta F=2$ is
\begin{eqnarray}
{\cal L}_{\rm eff}^{\Delta F=2} =\frac{G_F^2 M_W^2}{16\pi^2} 
\sum\limits_{a,b = u,c,t}^{} \lambda_a \lambda_b w_a w_b \left[S(w_a,w_b) +
X_aX_b \left\{2I_1(w_a,w_b,w_\xi)+X_aX_bI_2(w_a,w_b,w_\xi)\right\} \right] O_F \,.
\end{eqnarray}
Here, the $S(w_a,w_b)$ part is the SM contribution and the rest is due
to the charged Higgs box diagrams.  For $i$-type BGL model, $X_q =
-\cot\beta$ if $q=i$ and $X_q = \tan\beta$ otherwise.  The dimension-6
operator for $K^0$--$\bar{K}^0$ mixing is
\begin{equation}
O_F = (\bar{s}\gamma^\mu P_L d)^2 \,.
\end{equation}
Similar expressions can be obtained for $B$ systems. The relevant
parameters and functions are defined as follows:
\begin{eqnarray}
\lambda_a = V^*_{ad}V_{as} ~, ~~~~ w_a &=& \frac{m_a^2}{M_W^2} \,,
~~~f(x) = \frac{(x^2-8x+4)\ln x +3(x-1)}{(x-1)^2} \,,
\nonumber\\ S(w_a,w_b) &=& \frac{f(w_a)-f(w_b)}{w_a-w_b} \,,
~~~ g(x,y,z) = \frac{x(x-4)\ln x}{(x-1)(x-y)(x-z)} \,,
\nonumber\\ I_1(w_a,w_b,w_\xi) &=& [g(w_a,w_b,w_\xi) +
  g(w_b,w_\xi,w_a)+g(w_\xi,w_a,w_b)] \,,
\nonumber\\ I_2(w_a,w_b,w_\xi) &=&
\frac{1}{w_a-w_b}\left[\frac{w_a^2\ln w_a}{(w_\xi-w_a)^2}
  -\frac{w_b^2\ln w_b}{(w_\xi-w_b)^2} \right] \nonumber \\ && +
\frac{w_\xi [(w_\xi-w_a)(w_\xi-w_b)+\{2w_aw_b-w_\xi(w_a+w_b)\}\ln
    w_\xi]}{(w_\xi-w_a)^2(w_\xi-w_b)^2}\,.
\end{eqnarray}
Obtaining $M_{12}$ from the effective Lagrangian is
straightforward. As an example, for $K$-meson system (with $B_K$ as
bag parameter),
\begin{eqnarray}
&& M_{12}^K = -\frac{1}{2m_K} \Bra{K^0}{\mathscr L}_{\rm eff}^{\Delta
    F=2} \Ket{\bar{K^0}} \,, \\   {\rm with}~~~&& \Bra{K^0}O_F\Ket{\bar{K^0}} =
  \frac{2}{3}f_K^2 m_K^2 B_K \,.
\end{eqnarray}

\begin{center}
 \begin{figure}
 ~~\includegraphics[scale=0.45]{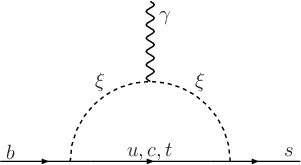}~~~~~
\includegraphics[scale=0.45]{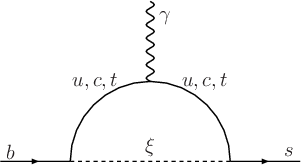}~~~~~
\includegraphics[scale=0.45]{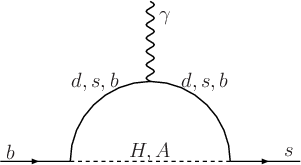}
\caption{\em Feynman diagrams involving nonstandard scalars
  contributing to $b\to s\gamma$ amplitude.}
\label{feyn_bsg}
\end{figure}
\end{center}

\section{Expressions for $b\to s\gamma$}
The effective Lagrangian for $b\to s\gamma$ can be written as
\begin{eqnarray}
{\cal L}_{\rm eff} &=& \sqrt{\frac{G_F^2}{8\pi^3}}V_{ts}^*V_{tb} m_b 
\left[\sqrt{\alpha}\left\{C_{7L} \bar{s}_L \sigma^{\mu\nu}b_R +
C_{7R} \bar{s}_R \sigma^{\mu\nu}b_L\right\}F_{\mu\nu}\right. \nonumber\\
&&~~~~~\left.\sqrt{\alpha_s} \left\{C_{8L} \bar{s}_L T_a \sigma^{\mu\nu}b_R +
C_{8R} \bar{s}_R T_a \sigma^{\mu\nu}b_L\right\}G^a_{\mu\nu}\right] +
\rm{h.c.} \, ,
\label{bsgam-effl}
\end{eqnarray}
where $F_{\mu\nu}$ and $G^a_{\mu\nu}$ are field strength tensors for photon
and gluon, respectively, and $T^a$s are the SU(3) generators. 
The branching ratio ${\rm Br}(b\to s\gamma)$ is given by 
\begin{eqnarray}
\frac{{\rm Br}(b\to s\gamma)}{{\rm Br}(b\to c e \bar{\nu})} = 
\frac{6\alpha}{\pi B}\left\vert 
\frac{V^*_{ts}V_{tb}}{V_{cb}}\right\vert^2 \left[  \left\vert C_{7L}^{\rm eff}\right\vert^2 
+\left\vert C_{7R}^{\rm eff}\right\vert^2 \right] \,,
\label{b2sgam}
\end{eqnarray}
where, we have taken $B=0.546$ \cite{Deschamps:2009rh}. 
The effective Wilson coefficients are
\begin{subequations}
\begin{eqnarray}
C_{7L}^{\rm eff} &=& \eta^{16/23} C_{7L} +
\frac{8}{3}\left(\eta^{14/23} -\eta^{16/23} \right) C_{8L} + {\cal C}
\,, \\ C_{7R}^{\rm eff} &=& \eta^{16/23} C_{7R} +
\frac{8}{3}\left(\eta^{14/23} -\eta^{16/23} \right) C_{8R} \,.
\end{eqnarray}
\label{effective Wilson}
\end{subequations}
In the above equations, $\eta = \alpha_s(M_Z)/\alpha_s(\mu)$, where
$\mu$ is the QCD renormalization scale; ${\cal C}$ corresponds to the
leading log QCD corrections in SM.  In the expression for the
effective Wilson co-efficients (Eq.\ (\ref{effective Wilson})), the
correction term is given by
\begin{eqnarray}
{\cal C} &=& \sum\limits_{i=1}^{8} h_i \eta^{a_i} \,,
\end{eqnarray}
where,
\begin{eqnarray}
a_i &=&
\left(\frac{14}{23},~\frac{16}{23},~\frac{6}{23},~-\frac{12}{23},~0.4086,~-0.4230,~-0.8994,~0.1456\right)\,,
\nonumber\\ h_i &=&
\left(\frac{626126}{272277},~-\frac{56281}{51730},~-\frac{3}{7},~-\frac{1}{14},~-0.6494,~-0.0380,~-0.0186,~-0.0057\right)
\,.
\end{eqnarray}
The values of $h_i$ and $a_i$ can be found in \cite{Gambino:2001ew}
[see Eq.\ (2.3) and Table 1 of Ref.~\cite{Gambino:2001ew}].

To understand the above expressions, we first define the following functions:
\begin{eqnarray}
\mathscr{F}_0(t) = \int\limits_{0}^{1} dx \frac{1-x}{x+(1-x)t} &=&
-\frac{1}{1-t}-\frac{\ln t}{(1-t)^2}\,, \nonumber\\ \mathscr{F}_1(t) =
\int\limits_{0}^{1} dx \frac{(1-x)^2}{x+(1-x)t} &=&
\frac{-3+4t-t^2}{2(1-t)^3}-\frac{\ln t}{(1-t)^3}\,,
\nonumber\\ \mathscr{F}_2(t) = \int\limits_{0}^{1} dx
\frac{(1-x)^3}{x+(1-x)t} &=& \frac{-11+18t-9t^2+2t^3-6\ln
  t}{6(1-t)^4}\,, \nonumber\\ \bar{\mathscr{F}}_0(t) =
\int\limits_{0}^{1} dx \frac{x}{x+(1-x)t} &=& \frac{1-t+ t\ln
  t}{(1-t)^2}\,,\nonumber \\ \bar{\mathscr{F}}_1(t) =
\int\limits_{0}^{1} dx \frac{x^2}{x+(1-x)t} &=&
\frac{1-4t+3t^2-2t^2\ln t}{2(1-t)^3}\,,
\nonumber\\ \bar{\mathscr{F}}_2(t) = \int\limits_{0}^{1} dx
\frac{x^3}{x+(1-x)t} &=& \frac{2-9t+18t^2-11 t^3+6t^3\ln
  t}{6(1-t)^4}\,.
\end{eqnarray}
Let us further define $x_t={m_t^2}/{m_W^2}$, $y_q={m_q^2}/{m_\xi^2}$,
$z_q={m_q^2}/{m_H^2}$, $z'_q={m_q^2}/{m_A^2}$. Now the
expressions for $C_{7 L},~C_{7 R},~C_{8 L},~C_{8 R}$ read
\begin{eqnarray}
C_{7L} &=& A_{\gamma}^{\rm SM} +A_{\gamma}^{\xi}+
\frac{Q_b}{V_{ts}^*V_{tb}}\sum\limits_{q=b,s}^{}\left[
  A_{L}^H(z_q)+A_{L}^A(z'_q)\right] \,, \nonumber\\ C_{7R} &=&
\frac{m_s}{m_b} A_{\gamma}^{\rm SM} + \frac{m_s}{m_b}A_{\gamma}^{\xi}
+
\frac{Q_b}{V_{ts}^*V_{tb}}\sum\limits_{q=b,s}^{}\left[A_{R}^H(z_q)+A_{R}^A(z'_q)\right]
\,,\nonumber \\
C_{8L} &=& A_{g}^{\rm SM} +A_{g}^{\xi}+\frac{1}{V_{ts}^*V_{tb}}\sum
\limits_{q=b,s}^{}\left[ A_{L}^H(z_q)+A_{L}^A(z'_q)\right] \,, \nonumber\\
C_{8R} &=& \frac{m_s}{m_b} A_{g}^{\rm SM} + \frac{m_s}{m_b}A_{g}^{\xi} +
\frac{1}{V_{ts}^*V_{tb}}\sum\limits_{q=b,s}^{}\left[A_{R}^H(z_q)+A_{R}^A(z'_q)\right] \,.
\end{eqnarray}
The SM and the new physics contributions (see Fig.~\ref{feyn_bsg}) are given below~:
\paragraph{$\blacksquare$ SM~:}
\begin{eqnarray}
A_{\gamma}^{\rm SM} &=&
\frac{1}{2}\left[\bar{\mathscr{F}}_1(x_t)+\bar{\mathscr{F}}_2(x_t) +
  \frac{1}{2}x_t\bar{\mathscr{F}}_2(x_t)
  -\frac{3}{2}x_t\bar{\mathscr{F}}_1(x_t) +
  x_t\bar{\mathscr{F}}_0(x_t) \right. \nonumber \\ &&~~~
  \left. +\frac{4}{3}\mathscr{F}_0(x_t) -2\mathscr{F}_1(x_t) +
  \frac{2}{3}\mathscr{F}_2(x_t)+\frac{1}{3}x_t\mathscr{F}_1(x_t)
  +\frac{1}{3}x_t\mathscr{F}_2(x_t) \right] -\frac{23}{36} \,,
\nonumber\\
A_{g}^{\rm SM} &=& \frac{1}{2} \left[2\mathscr{F}_0(x_t) -
3\mathscr{F}_1(x_t) +\mathscr{F}_2(x_t)+\frac{1}{2}x_t\mathscr{F}_1(x_t) +
\frac{1}{2}x_t\mathscr{F}_2(x_t)\right] - \frac{1}{3} \,, 
\end{eqnarray}
\paragraph{$\blacksquare$ Charged Higgs~:}
\begin{eqnarray}
A_{g}^{\xi} &=&
\frac{1}{4V_{ts}^*V_{tb}}\sum\limits_{q=u,c,t}^{}V_{qs}^*V_{qb}X_q^2
\left[y_q\mathscr{F}_1(y_q) + y_q\mathscr{F}_2(y_q) \right] \,,
\nonumber\\
A_{\gamma}^{\xi} &=&
\frac{1}{V_{ts}^*V_{tb}}\sum\limits_{q=u,c,t}^{}V_{qs}^*V_{qb}X_q^2
C(y_q) \,,
\end{eqnarray}
with 
\begin{equation}
 C(y) = \frac{1}{2}\left[\frac{1}{2}y\bar{\mathscr{F}}_2(y) 
 -\frac{3}{2}y\bar{\mathscr{F}}_1(y) +y\bar{\mathscr{F}}_0(y)
 +\frac{1}{3}y\mathscr{F}_1(y) +\frac{1}{3}y\mathscr{F}_2(y)\right] \,,
\end{equation}
and for $i$-type model, $X_q = -\cot\beta$ if $q=i$ and $X_q =
\tan\beta$ otherwise (e.g. for t-type model, $X_u=X_c=\tan\beta$,
$X_t=-\cot\beta$).
\paragraph{$\blacksquare$ CP-even Higgs~:}
\begin{eqnarray}
A_{L}^H(z_b) &=& -\frac{1}{8}\left[\left\{z_b\mathscr{F}_1(z_b) -
  z_b\mathscr{F}_2(z_b)\right\} \left(\frac{AD}{m_b^2} +
  \frac{BC}{m_b^2} \frac{m_s}{m_b} \right) +2z_b\mathscr{F}_1(z_b)
  \frac{AC}{m_b^2} \right] \,, \nonumber\\ A_{R}^H(z_b) &=&
-\frac{1}{8}\left[\left\{z_b\mathscr{F}_1(z_b) -
  z_b\mathscr{F}_2(z_b)\right\} \left(\frac{AD}{m_b^2} \frac{m_s}{m_b}
  + \frac{BC}{m_b^2} \right) +2z_b\mathscr{F}_1(z_b) \frac{BD}{m_b^2}
  \right] \,,
\end{eqnarray}
with $A=(N_d)_{sb} \,, B=(N_d)_{bs}^*\,, C=(N_d)_{bb}\,, D=(N_d)_{bb}^*\,.$ 
\begin{eqnarray}
A_{L}^H(z_s) &=& -\frac{1}{8}\left[\left\{z_s\mathscr{F}_1(z_s) -
  z_s\mathscr{F}_2(z_s)\right\} \left(\frac{AD}{m_s^2} +
  \frac{BC}{m_s^2} \frac{m_s}{m_b} \right) +2z_s\mathscr{F}_1(z_s)
  \frac{AC}{m_s^2} \frac{m_s}{m_b} \right] \,,
\nonumber\\ A_{R}^H(z_s) &=&
-\frac{1}{8}\left[\left\{z_s\mathscr{F}_1(z_s) -
  z_s\mathscr{F}_2(z_s)\right\} \left(\frac{AD}{m_s^2}\frac{m_s}{m_b}
  + \frac{BC}{m_s^2} \right) +2z_s\mathscr{F}_1(z_s) \frac{BD}{m_s^2}
  \frac{m_s}{m_b} \right] \,,
\end{eqnarray} 
with $A=(N_d)_{ss} \,, B=(N_d)_{ss}^*\,, C=(N_d)_{sb}\,, D=(N_d)_{bs}^*\,.$ 
\paragraph{$\blacksquare$ CP-odd Higgs~:}
\begin{eqnarray}
A_{L}^A(z'_b) &=& \frac{1}{8}\left[\left\{z'_b\mathscr{F}_1(z'_b) -
  z'_b\mathscr{F}_2(z'_b)\right\} \left(\frac{AD}{m_b^2} +
  \frac{BC}{m_b^2} \frac{m_s}{m_b} \right) +2z'_b\mathscr{F}_1(z'_b)
  \frac{AC}{m_b^2} \right] \,,\nonumber \\ A_{R}^A(z'_b) &=&
\frac{1}{8}\left[\left\{z'_b\mathscr{F}_1(z'_b) -
  z'_b\mathscr{F}_2(z'_b)\right\} \left(\frac{AD}{m_b^2}
  \frac{m_s}{m_b} + \frac{BC}{m_b^2} \right) +2z'_b\mathscr{F}_1(z'_b)
  \frac{BD}{m_b^2} \right] \,,
\end{eqnarray}
with $A=(N_d)_{sb} \,, B=-(N_d)_{bs}^*\,, C=(N_d)_{bb}\,, D=-(N_d)_{bb}^*\,.$
\begin{eqnarray}
A_{L}^A(z'_s) &=& \frac{1}{8}\left[\left\{z'_s\mathscr{F}_1(z'_s) -
  z'_s\mathscr{F}_2(z'_s)\right\} \left(\frac{AD}{m_s^2} +
  \frac{BC}{m_s^2} \frac{m_s}{m_b} \right) +2z'_s\mathscr{F}_1(z'_s)
  \frac{AC}{m_s^2} \frac{m_s}{m_b} \right] \,,
\nonumber\\ A_{R}^A(z'_s) &=&
\frac{1}{8}\left[\left\{z'_s\mathscr{F}_1(z'_s)
  -z'_s\mathscr{F}_2(z'_s)\right\}
  \left(\frac{AD}{m_s^2}\frac{m_s}{m_b} + \frac{BC}{m_s^2} \right) +
  2z'_s\mathscr{F}_1(z'_s) \frac{BD}{m_s^2} \frac{m_s}{m_b} \right]
\,,
\end{eqnarray}
with $A=(N_d)_{ss} \,, B=-(N_d)_{ss}^*\,, C=(N_d)_{sb}\,, D=-(N_d)_{bs}^*\,.$ 

\section{$B_s\to\mu^+\mu^-$}
The effective Hamiltonian is
\begin{eqnarray}
{\cal H}_{\rm eff} = C_{A}^{bs}O_{A}^{bs} + C_S^{bs}O_S^{bs} + C_P^{bs}O_P^{bs} \,, 
\end{eqnarray}
with
\begin{eqnarray}
O_{A}^{bs} = (\bar{b}\gamma_\alpha P_L s)(\bar{\mu}\gamma^\alpha
\gamma_5\mu) \,, ~~ O_{S}^{bs} = m_b(\bar{b} P_L
s)(\bar{\mu} \mu) \,, ~~ O_{P}^{bs} = m_b(\bar{b} P_L
s)(\bar{\mu} \gamma_5 \mu) \,.
\end{eqnarray}
Note that in addition to the above operators, there will be operators
of the form $(\bar{b}P_R s)(\bar{\mu}\mu)$ and $(\bar{b} P_R
s)(\bar{\mu} \gamma_5 \mu)$.  But the Wilson coefficients
corresponding to these operators will be proportional to $m_s$
(instead of $m_b$) and their contribution can be neglected ($m_b\gg
m_s$) as argued in \cite{Logan:2000iv}.  With this assumption we can
write
\begin{eqnarray}
\frac{{\rm Br}(B_s\to\mu^+\mu^-)}{{\rm Br}(B_s\to\mu^+\mu^-)_{\rm SM}}
&=& \left\{\left|1-m^2_{B_s}\frac{C_P^{bs}}{2m_\mu C_A^{bs}}\right|^2
+m_{B_s}^4\left(1-\frac{4m_\mu^2}{m_{B_s}^2}\right) \left|
\frac{C_S^{bs}}{2m_\mu C_A^{bs}}\right|^2 \right\} \times
\frac{\Gamma^{\rm SM}_B}{\Gamma_B}\,.
\end{eqnarray}
The relevant part of the Lagrangian to evaluate $C_S^{bs}$ and $C_P^{bs}$ is
\begin{eqnarray}
{\mathscr L}_{\rm quark} &=& \frac{R}{v} \bar{d}(N_dP_R + N_d^\dagger
P_L)d +i\frac{A}{v}\bar{d}(N_dP_R - N_d^\dagger P_L)d \nonumber \\ &=&
(N_d^\dagger)_{bs}\frac{R}{v} {\bar{b}P_Ls }
-i(N_d^\dagger)_{bs}\frac{A}{v} {\bar{b}P_Ls } \nonumber\\ &=&
(N_d^\dagger)_{bs}\frac{h}{v}\cos(\beta-\alpha) {\bar{b}P_Ls} -
(N_d^\dagger)_{bs}\frac{H}{v} \sin(\beta-\alpha){\bar{b}P_Ls } -
i(N_d^\dagger)_{bs}\frac{A}{v} {\bar{b}P_Ls }\,, \\ {\mathscr L}_{\rm
  lepton}&=& -\frac{H^0}{v}\bar{e}D_e e + \frac{R}{v} \bar{e}(N_eP_R +
N_e^\dagger P_L)e +i\frac{A}{v}\bar{e}(N_eP_R - N_e^\dagger P_L)e
\nonumber\\ &=& -\frac{m_\mu}{v}{ \bar{\mu}\mu H^0}
+\frac{(N_e)_{\mu\mu}}{v} {\bar{\mu}\mu R} + \frac{i(N_e)_{\mu\mu}}{v}
{\bar{\mu}\gamma_5\mu A} \nonumber\\ &=&\left[
  \frac{h}{v}\left\{-\sin(\beta-\alpha)m_\mu+\cos(\beta-\alpha)(N_e)_{\mu\mu}\right\}
  + \frac{H}{v}\left\{-\cos(\beta-\alpha)m_\mu -
  \sin(\beta-\alpha)(N_e)_{\mu\mu}\right\}\right] \bar{\mu}\mu
\nonumber \\ && +i \frac{A}{v}(N_e)_{\mu\mu}{\bar{\mu}\gamma_5\mu} \,.
\end{eqnarray}
Note that terms involving $\bar{b}P_Rs$ have not been displayed.
Their coefficients are proportional to $(N_d)_{bs}$, which is
proportional to $m_s$, and are therefore neglected.
\begin{subequations}
\begin{eqnarray}
(N_d)_{bs}&=& -(\tan\beta+\cot\beta)V^*_{ib}V_{is}m_s \,, \\
(N_d^\dagger)_{bs}&=& (N_d)^*_{sb}= -(\tan\beta+\cot\beta)V^*_{ib}V_{is}m_b \,.
\end{eqnarray}
\end{subequations}
The Wilson coefficients are
\begin{eqnarray}
C_S^{bs} &=& (\tan\beta+\cot\beta)\frac{V_{ib}^*V_{is}}{v^2}\left\{
\frac{\cos(\beta-\alpha)}{m_h^2}[-\sin(\beta-\alpha)m_\mu+\cos(\beta-\alpha)(N_e)_{\mu\mu}]
\right. \nonumber \\ && \left. ~~~~~~~~~~~~~~~~~~~~~~~~~~~~~
+\frac{\sin(\beta-\alpha)}{m_H^2}[\cos(\beta-\alpha)m_\mu+
  \sin(\beta-\alpha)(N_e)_{\mu\mu}]\right\}\,,
\end{eqnarray}
and
\begin{eqnarray}
C_P^{bs} &=& (\tan\beta+\cot\beta)\frac{V_{ib}^*V_{is}}{v^2}\frac{(N_e)_{\mu\mu}}{m_A^2} \,.
\end{eqnarray}
The SM Wilson coefficient is \cite{Logan:2000iv} 
\begin{equation}
C_A^{bs} =\frac{\alpha G_F}{2\sqrt{2}\pi \sin^2\theta_w} V_{tb}^*V_{ts} 2 Y(x_t)\,,\ \ \ 
Y(x_t)= 0.997 \left[\frac{m_t(m_t)}{166~{\rm GeV}}\right]^{1.55} \approx 1.0 \,. 
\end{equation}

\section{Leptonic and semileptonic decays}
The ratios $R(D)$ and $R(D^\ast)$ are defined as
\begin{equation}
R(D^{(*)}) = \frac{ {\rm Br}(B\to D^{(*)}\tau\nu)} 
{{\rm Br} (B\to D^{(*)}\ell\nu)}\,,
\end{equation}
where $\ell=e,\mu$.  The relevant expressions are \cite{rd}:
\begin{eqnarray}
\frac{R(D)}{R(D)_{\rm SM}} &=& 1 +1.5 {\rm Re}\left(\frac{C_R^{cb}+
C_L^{cb}}{C_{\rm SM}^{cb}}\right) 
+ 1.0 \left|\frac{C_R^{cb}+C_L^{cb}}{C_{\rm SM}^{cb}}\right|^2 \,, \nonumber\\
\frac{R(D^*)}{R(D^*)_{\rm SM}} &=& 1 +0.12 {\rm Re}\left(\frac{C_R^{cb}-C_L^{cb}}
{C_{\rm SM}^{cb}}\right) + 0.05 \left|\frac{C_R^{cb}-C_L^{cb}}{C_{\rm SM}^{cb}}
\right|^2 \,,\nonumber \\
\frac{{\rm Br}(B\to\tau\nu)}{{\rm Br}(B\to\tau\nu)_{\rm SM}} &=& \left|1+\frac{m_B^2}
{m_b m_\tau}\frac{(C_R^{ub}-C_L^{ub})}{C_{\rm SM}^{ub}}\right|^2 \,,
\end{eqnarray}
where we have assumed no appreciable change in the $B$-meson lifetime due to this
new interaction. The Wilson coefficients, as defined in the effective Hamiltonian in
Ref.~\cite{rd}, are
\begin{eqnarray}
 C_{\rm SM}^{qb} &=& 2\sqrt{2}G_F V_{qb}\,,\nonumber\\
 -C_R^{qb} &=& \frac{2}{v^2m_\xi^2} (VN_d)_{qb} (N_e^\dagger)_{\tau\tau} \,, \nonumber\\
-C_L^{qb} &=& -\frac{2}{v^2m_\xi^2} (N_u^\dagger V)_{qb} (N_e^\dagger)_{\tau\tau} \,,
\end{eqnarray}
where the extra minus sign in the last two lines comes from the nature of the propagator. 
For t-type model,
 \begin{eqnarray}
&& (N_e)_{\tau\tau} = -m_\tau \cot\beta \,, \nonumber\\
&& (N_u^\dagger V)_{ub} = m_u \tan\beta V_{ub} \,; ~~~~
(N_u^\dagger V)_{cb} = m_c\tan\beta V_{cb} \,, \nonumber\\
&& (VN_d)_{ub} = m_b\tan\beta V_{ub} \,; ~~~~ 
(VN_d)_{cb} = m_b \tan\beta V_{cb} \,.
\end{eqnarray}
Thus, none of the above decay widths depend on $\tan\beta$ for t-type model.

\end{document}